# Cognitive Agent Based Simulation Model For Improving Disaster Response Procedures

**Rohit K. Dubey**
D-INFK, ETH-Zurich
Future Cities Laboratory, Singapore-ETH Centre
Singapore
`rodubey@ethz.ch`

**Samuel S. Sohn**
Computer Science, Rutgers University
New Jersey, USA
`samuel.sohn@rutgers.edu`

**Christoph Hoelscher**
D-GESS, ETH-Zurich
Future Cities Laboratory, Singapore-ETH Centre
Singapore
`choelsch@ethz.ch`

**Mubbasir Kapadia**
Computer Science, Rutgers University
New Jersey, USA
`mubbasir.kapadia@rutgers.edu`

## Abstract

In the event of disaster, saving human lives is of utmost importance. For developing proper evacuation procedures and guidance systems, behavioural data on how people respond during panic and stress is crucial. In the absence of real human data on building evacuation, there is need for a crowd simulator to model egress and decision-making under uncertainty. In this paper, we propose an agent-based simulation tool, which is grounded in human cognition and decision-making, for evaluating and improving the effectiveness of building evacuation procedures and guidance systems during a disaster. Specifically, we propose a predictive agent-wayfinding framework based on information theory that is applied at intersections with variable route choices where it fuses $N$ dynamic information sources. The proposed framework can be used to visualize trajectories and prediction results (i.e., total evacuation time, number of people evacuated) for different combinations of reinforcing or contradicting information sources (i.e., signage, crowd flow, familiarity, and spatial layout). This tool can enable designers to recreate various disaster scenarios and generate simulation data for improving the evacuation procedures and existing guidance systems.

## 1 Introduction

The ability to evacuate people from densely populated, large, and complex buildings during a natural or human-made disaster is an essential design issue. To this end, engineers and designers rely on the conventional evacuation design codes and standards (e.g., door dimensions and the minimum number of exits) [1]. The analysis of recent disastrous incidents in buildings indicates that the conventional design codes and standards are not sufficient by themselves and highlights the need to study evacuation guidelines and the occupants' interactions with the building [2]. Cognitive agent-based simulation tools may aid the designers to (a) evaluate the evacuation procedures in both existing and future buildings and (b) highlight some of the problem areas (e.g., choke points during chaotic crowd motion and the lack of exit signs). Moreover, by simulating various unforeseen circumstances, such tools may aid in the training and decision-making strategies of first responders and building security personnel.

Wayfinding in a complex indoor environment is a dynamic process which is mentally demanding. Occupants have to continuously pick up relevant wayfinding cues from the environment, interpret them, and make route decisions accordingly. This process is dependent on physical (e.g., height, visual acuity) and psychological (e.g., attention, stress, anxiety, panic) factors. Stress and panic can influence navigational behaviour, which may differ vastly between an egress scenario and general circulation. Thus, there is a need to generate simulation data on how occupants might behave in the event of an emergency under panic, stress, chaotic crowd movement, and uncertainty. In the proposed cognitive agent-based framework, we decompose the directional decision into high and low levels, where a high-level decision (i.e., a *macro-decision*) chooses a global route to travel along, and a low-level decision (i.e., a *micro-decision*) chooses a local direction to move in. The wayfinding process becomes more pronounced when the occupant has to decide at an intersection. In this paper, we propose a simulation framework to model the macro- and micro-decision-making of agents at intersections based on previous research findings.

We specifically study the effects of four dynamic information sources under the influence of stress due to panic, but our framework generalizes to a variable number of sources: **Signage** - Occupants rely on signage in the absence of other wayfinding cues [3]. **Crowd flow** - Theoretical work has suggested that during egress and under stress, occupants may develop a tendency to follow others, a phenomenon called the "herding effect" [4, 5, 6]. **Spatial layout** - A corridor with longer radial line of sight [7, 8] and higher occlusivity [9] tends to bias human path choices. **Memory** - Familiarity to the environment along with reinforced information from other directional information such as signage or crowd flow influences human decision-making at a decision point. [10]. Moreover, when making a decision, an occupant will occasionally be confronted with conflicting information from different sources [11] (e.g., the person may receive a conflicting direction from a security guard compared to the direction provided by an "EXIT" sign during an evacuation). Therefore, it is essential to study the impact of these conflicts on a simulated agent's decision-making. To this end, we evaluate our proposed framework both under the dynamic change of an individual information source and in various combinations of either reinforcing (e.g., signage (**S**) + spatial layout (**P**), spatial layout (**P**) + crowd flow (**C**), etc.) or contradicting (e.g., **S** - **P**, **P** - **C**, **S** + **C** - **P**, etc.) information sources.

The proposed framework enables the facility manager/designer of a building to systematically evaluate the influence of environmental and psychological factors on egress performance in large and complex buildings. Moreover, the cognitively inspired decision-making model based on human uncertainty could help to improve the research in the field of disaster responses.

## 2 Related Work

Experiments and simulations which study crowd evacuation from a building during emergencies already exist. However, due to ethical and safety-related reasons, it is not possible to conduct an experiment with real participants. Therefore, in order to understand human behaviours, researchers typically focus on analysing previous events or develop computational models using human-like virtual agents. Here, we briefly mention some of the work done in agent-based egress models.

EvacSim models the egress of tall buildings with a large number of agents [12]. In this simulation, designers have the flexibility of selecting building behaviours by choice or by assigning a probability. Agents can interact with each other and with the environment. MASSEgress is an example of a pattern-based model wherein, agent behaviours are dependant on the surrounding environment, past experiences, and social or rational inferences [13]. One of the most successful pattern-based evacuation tools is buildingEXODUS [14]. Occupant, movement, behaviour, toxicity, and hazard are the five interacting elements which govern the simulation in buildingEXODUS. One of the drawbacks in pattern-based models is the predefinition of agent interactions which is computationally expensive and prohibits the modelling of unforeseen situations.

In the last two decades, many force-based egress models have been studied. Social force model [15] was the first one to study evacuee motion using a mixture of real (physical) and virtual (social) forces. Later on, many revisions of the social force model were implemented to improve its functionality [16, 17]. In the more recent past, force-based models have been criticised by a few researchers

**Limitations.** Some limitations of the works mentioned above are that they model the agent's decision-making either by predetermining their behaviour or by considering environment factors (e.g., signage, spatial layout, and crowd flow) in isolation. These models fail to both account for the absence of this isolation in varying degrees and generalize to new information. Bode et al. [10] show that while one-directional information cues (e.g., crowd flow and memory) may not affect occupants' decisions during an evacuation in isolation, they can have an influence when combined with other information sources.

# 3 Agent Wayfinding Prediction Model

In this section, we propose an agent wayfinding prediction model that fuses multiple information sources.

## 3.1 General Formalism

$$\mathbf{o}_t = O_t(l), \; \Gamma = \Gamma(l)$$
$$f_i(\mathbf{o}_t) = P_i(X \mid \mathbf{o}_t) \mid i = 1, \cdots, N-1$$
$$f_{mem}(\mathbf{o}_{\cdots}) = P_{mem}(X_t \mid \mathbf{o}_t, \mathbf{o}_{t-1}, \cdots, \mathbf{o}_{t-W+1})$$
$$\mathbf{F} = \left[ f_1(\mathbf{o}_t), \cdots, f_{N-1}(\mathbf{o}_t), f_{mem}(\mathbf{o}_t) \right]$$
$$G(\mathbf{F}) = P(x \mid \mathbf{F}) \mid x \in X$$

The preliminaries are defined as follows. $X$ is the set of all $M$ macro-decisions. Vector $\mathbf{o}_t$ consists of the observations made of the $N$ information sources at time $t$ from location $l$. $\Gamma$ is the set of neighboring positions for location $l$. The functions $f_i$ are constituent macro-decision-making models based on $N-1$ physical information sources and function $f_{mem}$ is a model based on memory as its information source. Matrix $\mathbf{F}_{M \times N}$ consists of the constituent models' probability distributions. Function $G$ fuses $\mathbf{F}$ into a single probability distribution over $X$.

Function $\Delta$ outputs the macro-decisions made by the information-theoretic framework, which thresholds the maximal macro-decision to determine whether to output it. Function $\delta$ evaluates the neighboring positions in $\Gamma$ and

$$\Delta(G(\mathbf{F})) = \begin{cases} arg\max_{x \in X}(G(\mathbf{F})) & max(G(\mathbf{F})) \geq \theta \\ \emptyset & max(G(\mathbf{F})) < \theta \end{cases}$$
$$\delta(N) = arg\max_{\gamma \in \Gamma} C\big(\mathbf{F}(O_t(\gamma))\big)$$

outputs the one (i.e., a micro-decision) where the information sources at that position maximize function $C$. At every time step $t$, $\mathbf{o}_t$ is updated and function $\delta$ is evaluated. Every $W$ time steps, function $\Delta$ is evaluated after having accumulated observations in memory.

## 3.2 Quantifying Directional Information from Multiple Sources

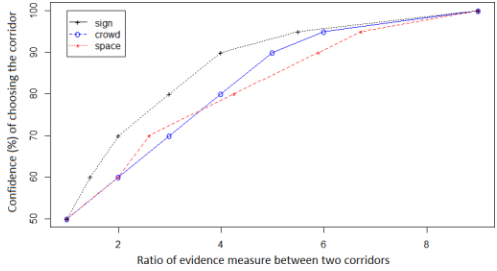

Figure 1: Three hypothetical probability distributions (Sign, Crowd, and Space) are employed to compute the confidence probability in choosing a corridor.

During the wayfinding decision process, an agent employs its perception model to quantify the information from $N$ sources (e.g., signage, spatial layout, crowd flow, and memory) that guides its macro-decisions in a virtual environment. The navigable areas are divided into an array of rectangular grid cells, which serve as reference points for an agent's location. The size of a grid cell is set to 0.5 meters by 0.5 meters because it approximates the average step length and size of an adult. Grid cell locations are pre-computed in the virtual environment and used as nodes in an 8-connected graph for agent navigation. The connectivity of this graph determines the function $\Gamma(l)$, which computes the neighboring cells at a given cell $l$ (Section 3.1). In order to realistically model the interaction between agents and the environment, a human-like visual perception model is used, which collects information at time $t$ into $\mathbf{o}_t$. The effective horizontal field of view (FOV) is 120 degrees in order to account for human neck rotation. Although the proposed framework supports $N$ information sources, we specifically use four information sources in the context of this paper to create decision-making models $f_i$.

**Signage.** We use an entropy-based information-theoretic principle to quantify the information provided by a sign in a virtual environment. In particular, a function is used to compute the visibility information from a sign $s \in \mathbf{o}_t$ at location $l$. The visibility-based confidence in a macro-decision is proportional to the relative angle and the distance between sign $s$ and location $l$ [3]. An entropy-based decision confidence distribution (Figure 1) is then employed to compute the confidence probability $f_{sign}(\mathbf{o}_t)$ of perceiving the directional information for each possible macro-decision $x \in X$.

**Spatial Layout.** To quantify space, we rely on four isovist measures (i.e., max radial line, isovist area, isovist perimeter, and isovist occlusivity) [18]. Instead of using one single isovist per intersection based on the agent's FOV, we divide it into $M$-many (i.e., the number of possible macro-decisions $X$) smaller partial isovists. The aforementioned isovist measures are computed for each partial isovist based on [18] and then aggregated. To compute the confidence probability $f_{space}(O)$ afforded to each macro-decision, we employ a hypothetical distribution (Figure 1) based on the ratio of the partial isovist measures [8, 7, 19].

**Crowd Flow.** To compute an agent's perception of crowd flow for each macro-decision, we have a function $f_{crowd}$ compute the number of visible agents inside the isovist polygon for that route. Then, a hypothetical distribution (Figure 1) is used to generate the confidence probability $f_{crowd}(\mathbf{o}_t)$ for each macro-decision.

**Memory.** The memory information source applies the function $f_{mem}$ to fuse the probability distributions of the $N-1$ other information sources over $W = 3$ time steps using observations $\mathbf{o}_{t-W+1}, \ldots, \mathbf{o}_t$. The probability distributions are first converted to beliefs and then combined using a temporal weighted combination rule [20], which effectively weights time steps to value newer information over older information.

### 3.3 Information-Theoretical Framework

The framework described in this section takes as input $N$ probability distributions in the form of $\mathbf{F}$ and outputs either one of $M$ macro-decisions or no decision. Based on [21], a multi-source information fusion method is proposed that considers Jensen-Shannon divergence (JSD) and Shannon entropy ($H$) in order to determine the confidence in each of the macro-decisions. JSD is employed to measure uncertainty between information sources and entropy is used to measure uncertainty within information sources. The steps involved in this information-theoretical approach are described below.

**Step 1:** We compute the JSD between each pair of sources, where $A = \frac{2 \cdot \mathbf{F}_i(x)}{\mathbf{F}_i(x) + \mathbf{F}_j(x)}$ and $B = \frac{2 \cdot \mathbf{F}_j(x)}{\mathbf{F}_i(x) + \mathbf{F}_j(x)}$.

$$JSD_{i,j} = \frac{1}{2} \sum_x^X \mathbf{F}_i(x) \cdot log_2(A) \quad (1)$$
$$+ \frac{1}{2} \sum_x^X \mathbf{F}_j(x) \cdot log_2(B)$$

**Step 2:** The average JSD ($JSD_i^\mu$) of information source $i$ can be calculated by Equation 2.

$$JSD_i^\mu = \frac{\sum_{j=1, j \neq i}^N JSD_{i,j}}{N-1} \quad (2)$$

**Step 3:** The support degree $Sup_i$ of information source $i$ is defined in Equation 3, where $\epsilon = 10^{-5}$ is used in practice.

$$Sup_i = \frac{1}{max(\epsilon, JSD_i^\mu)} \mid 0 < \epsilon \ll 1 \quad (3)$$

**Step 4:** The credibility degree $Crd_i$ of information source $i$ is defined in Equation 4, where the range of $Crd_i$ is $[0, 1]$.

$$Crd_i = \frac{Sup_i}{\sum_{j=1}^N Sup_j} \cdot \left(1 - \frac{1}{N} \sum_{j=1}^N JSD_j^\mu\right) \quad (4)$$

**Step 5:** We then measure the normalized Shannon entropy of each information source $i$, which is the entropy of source $i$ divided by the maximum possible entropy for $X$.

$$\tilde{H}_i = -\sum_x^X \mathbf{F}_i(x) \cdot log_2 \mathbf{F}_i(x) / log_2 \frac{1}{M} \quad (5)$$

**Step 6:** Based on the normalized entropy $\tilde{H}_i$, the credibility degree $Crd_i$ is adjusted, giving the confidence in the information provided by source $i$. On account of the confidence in each information source $i$, the confidence distribution over $X$ will be obtained by Equation 6.

$$G = \sum_{i=1}^N Crd_i \cdot (1 - \tilde{H}_i) \cdot \mathbf{F}_i \quad (6)$$

Steps 1 through 6 correspond to $G(\mathbf{F})$ (Section 3.1), which transforms the input $\mathbf{F}$ into the output confidence distribution over $X$. The rule that this framework uses to output a macro-decision is based on the highest confidence in the output distribution. If this value exceeds $\theta$, the framework is sufficiently confident in the corresponding macro-decision, which it makes. Otherwise, the framework will not make a decision. This decision rule corresponds to function $\Delta(G(\mathbf{F}))$.

### 3.4 Agent Decision Model

According to [22], agents should take three steps between making macro-decisions in order to simulate realistic wayfinding. The memory information source is the only source that spans multiple time steps, and it accommodates this type of decision-making by ensuring that no time step has its observations ignored. However, without changing the location of the agent, the memory will not be fusing different probability distributions per each information sources. Therefore, we must have the agent move while it is deliberating on its macro-decision by making one micro-decision $\delta$ per time step (Section 3.1). This micro-decision is either a $\frac{1}{2}$ meter or $\frac{2}{2}$ meter step in the direction $\gamma$ that maximizes function $C$, which takes the maximum probability of a macro-decision given by either signage $f_{sign}(O_t(\gamma))$ or spatial layout $f_{space}(O_t(\gamma))$. This micro- and macro-decision-making cycle repeats until a physical user-defined threshold is reached or the agent within a certain proximity to the intersection. In either case, the final macro-decision made by function $\Delta$ is chosen as the agent's goal direction, while all prior macro-decisions are predictions of the agent's goal direction at that point in time.

## 4 Experiments & Results

This section describes the simulation performed to verify the proposed dynamic, uncertain information fusion framework. The general test-case we present is a wayfinding decision-making problem at an intersection/decision point with two and four route choices under the influence of multiple information sources.

### 4.1 Effects of Reinforced and Contradictory Combinations of Multiple Information Sources

100 agents were spawned randomly and assigned a wayfinding task of finding a target location (e.g., Find Gate A2) for each test-case. In Table 1, we presents the prediction results for eight test-cases.

| Test Cases | Left (**L**) | | | Right (**R**) | | | Prediction Results | |
|---|---|---|---|---|---|---|---|---|
| | **S** | **C** | **P** | **S** | **C** | **P** | Information Theory | |
| 1 ($S^+$, $C^+$, $P^+$) | Yes | High | High | No | Med | Med | **L** (89%) | **R** (11%) |
| 2 ($S^+$, $C^-$, $P^-$) | Yes | Low | Low | No | High | High | **L** (4%) | **R** (96%) |
| 3 ($S^+$, $C^-$, $P^-$) | Yes | Med | Low | No | High | High | **L** (5%) | **R** (95%) |
| 4 ($C^-$, $P^+$) | No | Low | Med | No | Med | Med | **L** (5%) | **R** (95%) |
| 5 ($C^+$, $P^-$) | No | Low | High | No | Med | Low | **L** (6%) | **R** (94%) |
| 6 ($C^-$, $P^+$) | No | Low | High | No | Low | Low | **L** (27%) | **R** (73%) |
| 7 ($C^-$, $P^+$) | No | High | Med | No | Low | High | **L** (44%) | **R** (56%) |
| 8 ($C^+$, $P^-$) | No | High | Low | No | Med | High | **L** (11%) | **R** (89%) |

Table 1: Prediction results from the proposed method are shown in various combinations of information sources (Signage **S**, Crowd Flow **C**, and Spatial Layout **P**).

**Sign+Crowd+Space ($S^+$, $C^+$, $P^+$)**: Test-case 1 represents the reinforced combination. All three information sources are strongly affording the directional information to take "Left". We found that almost 90% of the agents predicted to go left. It is in agreement with previous work stating that the presence of sign has a strong effect on human wayfinding decision making.

**Sign - (Crowd+Space) ($S^+$, $C^-$, $P^-$)**: Test-cases 2 and 3 represent the conflicting information from the sign and reinforced information from the combination of crowd and space. In both test-cases, almost all agents' decisions were influenced by the strong presence of the crowd and space.

**Crowd - Space ($S^+$, $S^-$)**: Test-cases 4 to 8 represent the effect of conflicting information from crowd and space in the absence of a sign. By decreasing the confidence from the crowd's presence (for right corridor) in test-case 6, we observe a considerable reduction in the number of agents selecting that corridor in comparison to test-case 5.

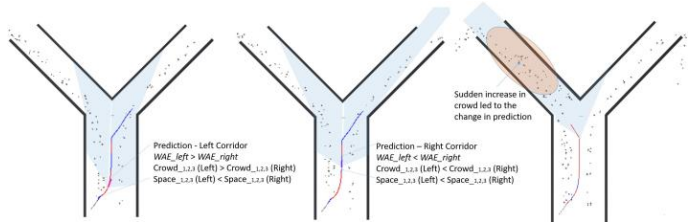

Figure 2: We highlight two sections in the agent decision-making process separated over time. In the beginning phase, agents predict to take the left route due to its position and orientation. Agents perceive higher confidence from the crowd on the left corridor, and the difference in the confidence between the left and right corridors afforded by space is small. In the later phase (middle image), agent perception of both crowd and space changes and the confidence afforded by space and crowd gradually increases towards the right corridor. We replicate the same set-up and increase the crowd flow at the far end of the left corridor (rightmost image). The agent is exposed to this new crowd only at the later phase of decision-making, resulting in the change of its prediction.

The results observed from the proposed framework highlights its non-deterministic nature. We believe this is due to two main reasons. Firstly, the direction of approach of an agent at an intersection influences the confidence afforded by various information sources and secondly, the continuous change in the information over a temporal axis.

## 5 Conclusions

In the absence of real crowd behaviour data in disasters and emergencies, a simulation tool which mimics human decision making capability and models the crowd emergence behaviour due to micro-level occupant behaviour is a viable alternative. In this paper, we have proposed an information-theoretic agent wayfinding prediction framework which predicts an agent's navigational decision at an exit/intersection with $M$ route choices under the influence of $N$ information sources. The simulation results highlight the non-deterministic nature of our framework and produces realistic results that are consistent with previous works. We demonstrate that our information-theoretic method is also able to fuse the uncertain and dynamically changing information over time. The proposed work can be used to model the individual-level interactions and decision-making of an agent and can be used to study the evacuation behaviour or general circulation of a crowd in an indoor environment. One limitation of our proposed method is its dependence on probability distributions assigned to individual information sources. A small error in the distribution can result in faulty decisions. Also, the mutual information between $N$ information source is not considered. We aim to extend our model by utilizing context as an information source of information as proposed in [23].

# 6 Appendix

## 6.1 Influence of Memory

In the proposed temporal weighted evidence combination model we encode the memory of an agent as a fourth information source. We model memory as the continuous process of information retention over time. The results generated in this paper (see Table 1) is by considering memory for $W = 3$ time steps (i.e., information from past three time steps is fused according to their temporal weight during decision-making). Our hypothesis is that the uncertainty in an agent's prediction will reduce with the increase in memory time steps. In Figure 3 we plot the impact of the increase in memory duration $W$ on the agent's prediction entropy. We notice a gradual reduction in prediction entropy with the increase in memory. How long the memory should be retained in the context of wayfinding decision-making needs a thorough examination which is beyond the scope of this paper. The result demonstrates that our framework is capable of modelling working memory in an agent wayfinding prediction model.

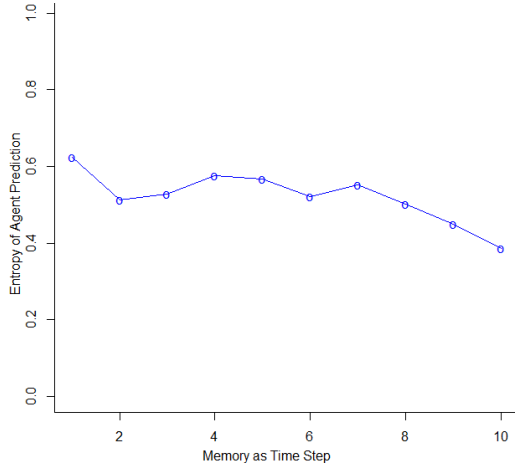

Figure 3: Influence of memory duration $W$ on agent's prediction uncertainty.

## 6.2 Generalization of The Proposed Model

In Figure 4 we expand the number of possible route choices from two to four to showcase the generalization capability of the proposed model to $M$ number of possible route choices. We present four different examples under various combination of information sources. In the first example (Fig. 4(a)), we only keep the impact of spatial layout. The highest confidence of spatial layout evidence is highlighted in light orange (Route 2). Unsurprisingly, the agent decides to take Route 2. In Figure 4(b), we remove the influence of signage and our model adapts to $N = 2$ information sources (i.e., crowd flow and spatial layout). Different prediction results at different stages of agent wayfinding are color-coded. We believe, the change in prediction at different time intervals models the fluctuation in the human decisions during a wayfinding task due to the increase or decrease in evidences' confidences. The model successfully captures the temporal property of information sources. In Figure 4(c), we introduce a directional sign which directs to Route 1. In Figure 4(d), we remove the sign and allow the agent to get influenced by the fused information sources (i.e., **C** and **P**). Similarly to Figure 4(b), the prediction fluctuates over time and finally, the agent decides to choose Route 3.

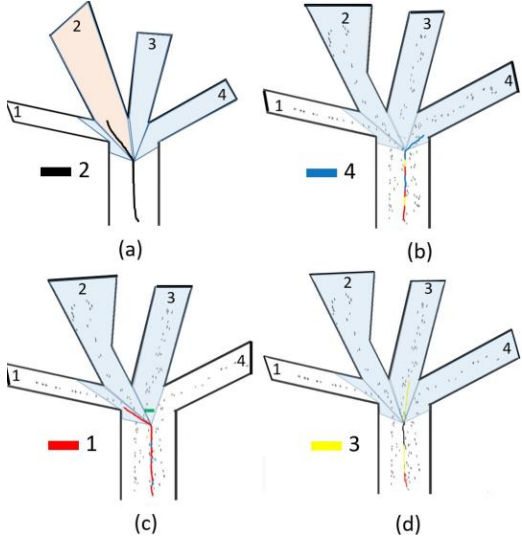

Figure 4: Visualization of an agent prediction model at an intersection with four route choices. Representation of the proportion of crowd in each corridor is shown in black dots. Isovist polygon for each corridor is shown in a light blue semi-transparent polygon. Agent's trajectory is colour-coded to represent its prediction state in that time.